\documentclass[aps,pre,twocolumn,showpacs,superscriptaddress,groupedaddress,notitlepage]{revtex4-1}
\usepackage[english]{babel}
\usepackage{amsmath,amssymb}
\usepackage{graphicx}
\usepackage{dcolumn}
\usepackage{bm}
\usepackage{amsmath}
\usepackage{xcolor}
\usepackage[applemac]{inputenc}
\usepackage{mathtools}
\usepackage{accents}
\usepackage[labelformat=empty,caption=false]{subfig}
\usepackage{keyval}
\usepackage{bm}
\newlength{\dhatheight}

\DeclareMathOperator*{\eq}{=}
\newcommand{\Average}[1]{\left\langle #1 \right\rangle}

\newcommand*{\diff}{\mathop{}\!\mathrm{d}}
\renewcommand{\phi}{\varphi}

\newcommand{\derpar}[2]{\frac{\partial #1}{\partial #2}}

\newcommand{\dersecpar}[2]{\frac{\partial^2 #1}{\partial #2^2}}

\newcommand{\fracder}[2]{\mathcal{D}_{#1}^{#2}}
\newcommand{\MittagLeffler}[3]{E_{#1,#2}(#3)}

\newcommand{\LaplaceInverse}[1]{\mathcal{L}^{-1} \left\{#1\right\}}

\addto{\captionsenglish}{}

\begin{document}

\hspace{5.2in} \mbox{Fermilab-Pub-04/xxx-E}

\title[]{Langevin formulation of a subdiffusive continuous time random walk in physical time}
\author{Andrea Cairoli}
\affiliation{School of Mathematical Sciences, Queen Mary, University of London, Mile End Road, E1 4NS, UK}
\author{Adrian Baule}
\email{Correspondence to: a.baule@qmul.ac.uk}
\affiliation{School of Mathematical Sciences, Queen Mary, University of London, Mile End Road, E1 4NS, UK}
\date{\today}
\begin{abstract}
Systems living in complex non equilibrated environments often exhibit subdiffusion characterized by a sublinear power-law scaling of the mean square displacement. One of the most common models to describe such subdiffusive dynamics is the continuous time random walk (CTRW). Stochastic trajectories of a CTRW can be described in terms of the subordination of a normal diffusive process by an inverse L{\'e}vy-stable process. Here, we propose an equivalent Langevin formulation of a force-free CTRW without subordination. By introducing a new type of non-Gaussian noise, we are able to express the CTRW dynamics in terms of a single Langevin equation in physical time with additive noise. We derive the full multi-point statistics of this noise and compare it with the scaled Brownian motion (SBM), an alternative stochastic model describing subdiffusive dynamics. Interestingly, these two noises are identical up to the 2nd order correlation functions, but different in the higher order statistics. We extend our formalism to general waiting times distributions and force fields and compare our results with those of SBM. In the presence of external forces, our proposed noise generates a new class of stochastic processes, resembling a CTRW but with forces acting at all times.
\end{abstract}


\maketitle

\section{\label{Sec:0}\textbf{INTRODUCTION}}

Many systems in nature live in complex non-equilibrated or highly crowded environments, thus exhibiting anomalous diffusive patterns, which deviate from the well known Fick's law of purely thermalized systems \cite{klafter1996beyond,metzler2000random,sokolov2012models}. Their distinctive feature is the power-law scaling of the mean-square displacement (MSD) \cite{shlesinger1995levy,klafter1996beyond,pekalski1999anomalous,metzler2000random,sokolov2012models}:
\begin{equation}
\Average{(Y(t)-Y_0)^2}\sim t^{\alpha},
\label{eq:0.1}
\end{equation}        
where $\Average{\,\cdot\,}$ indicates the ensemble average over different realizations of the stochastic process $Y(t)$ describing the dynamics, usually either a velocity or a position, $Y_0$ is its initial condition and $\alpha \in \mathbb{R}^{+}$. While Fick's law is recovered by setting $\alpha=1$, thus predicting for normal diffusion the typical liner scaling of the MSD, we can distinguish between different types of anomalous behaviour. Indeed, we define subdiffusion if $0<\alpha<1$ and superdiffusion if $\alpha>1$, which correspond to processes dispersing with a slower or faster pace than Brownian motion respectively. 

Examples of such anomalous processes were first found in physical systems, such as charge carriers moving in amorphous semiconductors, particles being transported on fractal geometries or diffusing in turbulent fluids/plasma or in heterogeneous rocks (see \cite{metzler2000random} and references therein). However, with the recent improvements of experimental techniques in biology, joint position-velocity datasets have been obtained, which are revealing the existence of many more examples in living systems. On the one hand, cell migration experiments have revealed a characteristic superdiffusive scaling of the MSD along with many more features deviating from standard Brownian models, e.g., non Gaussian probability density functions (PDFs) for the position and/or the velocity of the moving cell and power-law long time scaling of the velocity auto-correlation functions (see Refs.~\cite{selmeczi2005cell,*selmeczi2008cell,*dieterich2008anomalous,*reynolds2010can,*campos2010persistent,*harris2012generalized} and references therein). On the other hand, biological macromolecules and/or organelles often exhibit a subdiffusive scaling of the MSD, while moving in molecular crowded environments. These can be prepared ad hoc for in vitro experiments, e.g., by using solutions of surfactant micelles \cite{szymanski2006movement} or polymer networks \cite{wong2004anomalous} to name just a few techniques, or found in vivo, e.g., in the cytoplasm \cite{weber2010bacterial} or the cells' membrane \cite{weigel2011ergodic}, whose viscoelastic properties have recently been found to play a major role in determining the anomalous diffusion \cite{szymanski2009elucidating,Pan2009Viscoelasticity}. Furthermore, biological systems may also exhibit a rich dynamical behavior, such as non linear MSDs, showing crossover between different scaling regimes \cite{levi2005chromatin,brangwynne2007force,bronstein2009transient,xu2011subdiffusion,jeon2011vivo,tabei2013intracellular,javer2013short,javer2014persistent} at different timescales, and/or a dependence of the corresponding diffusion coefficients on energy-driven active mechanisms \cite{levi2005chromatin,weber2012nonthermal,*mackintosh2012active,javer2013short,*javer2014persistent}. We refer the interested reader to \cite{hofling2013anomalous} for a recent review on anomalous diffusive systems. 

Considering this wide, though not exhaustive, variety of different anomalous behaviours, one needs to have a tool-kit of well studied models with which trying to fit the experimental data and infer the specific microscopic processes underlying the observed dynamics. Here we focus on subdiffusive processes, for which many models have been introduced so far, which are capable of reproducing the characteristic scaling of Eq.~\eqref{eq:0.1}, while still showing distinct features if we look at other properties, like the multipoint correlation functions \cite{baule2005joint,vsanda2005multipoint,baule2007two,*baule2007fractional,barkai2007multi,*niemann2008joint,*meerschaert2012fractional}. Among the most commonly applied to data analysis, we find the continuous-time random walk (CTRW) \cite{montroll1965random,metzler2000random} and the scaled Brownian motion (SBM) \cite{lim2002self,wu2008propagators,thiel2014scaled,Jeon2014ScaledBM}. 

In the seminal paper \cite{montroll1965random}, the CTRW was introduced as a natural generalization of a random walk on a lattice, with waiting times between the jumps and their size being sampled from general and independent probability distributions. Only later, a convenient stochastic representation of these processes was derived in terms of subordinated Langevin equations \cite{fogedby1994langevin}, which provided a suitable formalism to derive their multipoint correlation functions \cite{baule2005joint,baule2007fractional,baule2007two}. Although the focus was first on power-law distributed waiting times, which indeed provided Eq.~\eqref{eq:0.1} exactly for all times, recent works adopted more general distributions \cite{chechkin2008generalized,*chechkin2002retarding,*magdziarz2009langevin,stanislavsky2014anomalous,Cairoli2014anomalous}, thus being able to model the crossover phenomena so often occurring in biological experiments.

On the other hand, the SBM has been recently introduced as a Gaussian model of anomalous dynamics \cite{lim2002self}, providing the same scaling of Eq.~\eqref{eq:0.1} for all its dynamical evolution. If $B(t)$ is a usual BM, its scaled version is defined by making a power-law change of time with exponent $\alpha$: $B(t^{\alpha})$. Although being commonly used to fit data \cite{mitra1992diffusion,szymanski2006movement,wu2008propagators}, it has recently been shown to be a non stationary process with paradoxical behaviour under confinement, i.e., in the presence of a linear viscous-like force, as the corresponding MSD unboundedly decreases towards zero. This is suggested to be ultimately caused by the time dependence of the environment, either of the temperature or of the viscosity. As a consequence, it has been ruled out as a possible alternative model of anomalous thermalized processes \cite{Jeon2014ScaledBM}.  

In this paper, we derive a new type of noise, which allows us to express a free diffusive CTRW in terms of a single Langevin equation in physical time. We provide the full characterization of its multipoint correlation functions and we compare them with those of the noise driving a SBM. Both purely power-law waiting times and general waiting time distributions are discussed \cite{stanislavsky2014anomalous,Cairoli2014anomalous}. We find that the correlation functions are identical up to the two point ones, but different for higher orders: the noise driving SBM is a Gaussian noise, while our new noise driving a CTRW is clearly non-Gaussian. Here, all odd correlation functions vanish, as for Gaussian noise, but the even ones do not satisfy Wick's theorem \cite{isserlis1918formula,*wick1950evaluation}. The newly defined noise enables us to define a class of CTRW like processes with forces acting for all times, which are different from the corresponding standard CTRWs. Furthermore, we revisit the behaviour of the SBM under confinement and show that its MSD correctly converges to a plateau as it is typical of confined motion \cite{burov2011single}, provided that we use more general time changes with truncated power-law tails. This suggest that the anomaly observed in \cite{Jeon2014ScaledBM} is mainly due to the localizing effect of the external linear force, which is able to trap the particle in the zero position if we allow for infinitely long waiting times between the jumps to eventually occur in the long time limit.

\section{\label{Sec:1}\textbf{CTRWS, scaled BM and generalization to arbitrary waiting times' distributions and time transformations}}
We recall in this first section definitions and properties of the free diffusive CTRW and SBM, which will be useful later in the discussion. We are mainly interested in their stochastic Langevin formulation and in both their Fokker-Planck (FP) equation and MSD. We then generalize these results to the case of arbitrary waiting times' distributions or time transformations for CTRW and SBM respectively. 

Throughout the discussion, $\widetilde{f}$ indicates the Laplace transform of a function $f(t)$ defined on the positive half line: $\widetilde{f}(\lambda)=\mathcal{L}\left\{f(t)\right\}(\lambda)=\int_0^{+\infty}e^{- \lambda t}f(t)\diff{t}$. Moreover, $\phi_1 * \phi_2$ denotes the convolution of two functions $\phi_1$ and $\phi_2$, defined on the positive half line: $(\phi_1 * \phi_2)(t)=\int_{0}^{t} \phi_1(t-\tau) \phi_2(\tau) \diff{\tau}$. We remark that the corresponding definitions for functions of multiple variables follow straightforwardly. 

\subsection{\label{SubSec:1a}\textbf{CTRW}}
A Langevin representation of a CTRW was first proposed in \cite{fogedby1994langevin}, where the method of the stochastic time-change of a continuous-time process is used. Its set-up consists in introducing two auxiliary processes $X(s)$ and $T(s)$, which we assume for now to be purely diffusive and L\'evy stable with parameter $\alpha$ ($0<\alpha\leq 1$) respectively. They both depend on the arbitrary continuous parameter $s$ and have dynamics described in terms of Langevin equations \cite{fogedby1994langevin}: 
\begin{subequations}
\begin{align}
\dot{X}(s)&=\sqrt{2 \sigma }\,\xi(s)\label{eq:1.1a}\\
\dot{T}(s)&=\eta(s) \label{eq:1.1b}
\end{align}
\end{subequations}
where $\xi(s)$ and $\eta(s)$ are two independent noises. For $X(s)$ to be a normal diffusion, we require $\xi(s)$ to be a white Gaussian noise with $\Average{\xi(s)}=0$ and $\Average{\xi(s_1)\xi(s_2)}=\delta(s_2-s_1)$. On the other hand, $\eta(s)$ is a stable L\'evy noise with parameter $\alpha$ ($0<\alpha\leq 1$) \cite{cont1975financial}. The anomalous CTRW is then derived by making a randomization of time, i.e., by considering the time-changed (or subordinated) process: $Y(t)=X(S(t))$, with $S(t)$ being the inverse of $T(s)$, defined as a collection of first passage times:
\begin{equation}
S(t)=\inf_{s>0}{\left\{s:T(s)>t\right\}}.
\label{eq:1.2}
\end{equation}
The process $Y(t)$ is easily shown to satisfy Eq.~\eqref{eq:0.1} exactly for all its time evolution, by recalling that the probability density function (PDF) of $S(t)$ has the Laplace transform $\widetilde{h}(s,\lambda)=\lambda^{\alpha-1} e^{-s\lambda^{\alpha}}$ \cite{baule2005joint} and that $\Average{X^2(s)}=2\,\sigma\,s$. Indeed, we obtain in Laplace space:
\begin{equation}
\Average{\widetilde{Y}^2(\lambda)}=\int_0^{+\infty}\Average{X^2(s)}\widetilde{h}(s,\lambda)\diff{s}=\frac{2\,\sigma}{\lambda^{1+\alpha}},
\label{eq:1.3}
\end{equation}     
whose inverse transform confirms its anomalous scaling: 
\begin{equation}
\Average{Y^2(t)}=\frac{2\,\sigma}{\Gamma(1+\alpha)}t^{\alpha}. 
\label{eq:1.4}
\end{equation}
As expected, this same MSD is obtained by taking the diffusive limit of the microscopic random walk formulation of the CTRW, where we allow for asymptotically power-law distributed waiting times between the jumps of the walker, whose sizes are drawn from a distribution with finite variance \cite{metzler2000random}. In this limit, the model also provides a fractional diffusion equation for the PDF of $Y(t)$:
\begin{equation}
\derpar{}{t}P(y,t)=D_{\alpha}\dersecpar{}{y}\fracder{t}{1-\alpha}P(y,t), 
\label{eq:1.5}
\end{equation}                     
where $D_{\alpha}$ is a generalized diffusion coefficient and $\fracder{t}{1-\alpha}f(t)=\frac{1}{\Gamma(\alpha)}\derpar{}{t}\int_0^{t}\left(t-\tau\right)^{\alpha-1}f(\tau)$ is the Riemann-Liouville time-integral operator, which makes the non Markovian character of the CTRW evident. It is then natural to investigate if the set of Eqs.~(\ref{eq:1.1a}-\ref{eq:1.1b}) can give this same FP equation. This has been proved in \cite{magdziarz2008equivalence,eule2012langevin,Cairoli2014anomalous}, with the specification: $D_{\alpha}=\frac{\sigma}{\Gamma\left(1+\alpha\right)}$, thus confirming the equivalence in the diffusive limit of the original random walk model and of the subordinated Langevin Eqs.~(\ref{eq:1.1a}-\ref{eq:1.1b}).  

We remark that the formulation of CTRWs as a subordinated normal diffusive processes can be considered as the continuum limit of the original renewal picture of Montroll and Weiss \cite{montroll1965random}. Here, the position $Y(t)$ of a CTRW is characterized by two sets of random variables $\{(\xi_i,\eta_i)\}_{i,\ldots,N(t)}$, with $N(t)$ being the number of jumps up to the time $t$. The random variables $\xi_i$ and $\eta_i$ specify the amplitude of the jumps occurring at the random time $t_i$, i.e., $\xi_i=Y(t_i)-Y(t_{i-1})$, and the waiting times between two successive jumps, i.e., $\eta_i=t_i-t_{i-1}$. Thus, $Y(t)$ is obtained by summing all the variables $\xi_i$:
\begin{equation}
Y(t)=\sum_{i=1}^{N(t)}\xi_i.
\label{eq:add1}
\end{equation}
In the present discussion, we assume $\{\xi_i\}$ and $\{\eta_i\}$ separately to be i.i.d random variables and each $\xi_i$ to be independent of $\eta_i$. On the other hand, we obtain by direct integration of Eqs.~(\ref{eq:1.1a}-\ref{eq:1.1b})
\begin{align}
Y(t)&=X(S(t))\notag\\
&=\int_0^{S(t)}\dot{X}(\tau)\diff{\tau}\notag\\
&=\sqrt{2\,\sigma}\int_0^{S(t)} \xi(\tau) \diff{\tau}.
\label{eq:add2}
\end{align}
Therefore, Fogedby's approach \cite{fogedby1994langevin} describes the resulting trajectory of the random walk in the continuum limit by parametrizing both the path of the walker $X(\cdot)$ and the time elapsed $T(\cdot)$ with an arbitrary continuous arc-length $s$. The stochastic process $S(\cdot)$ is the inverse of $T(\cdot)$ and measures the arc-length as a function of the physical time. $S(t)$ thus represents the continuum limit of the random variable $N(t)$ that counts the number of steps in the renewal picture.

\subsection{\label{SubSec:1b}\textbf{SCALED BM}}

If instead of a stochastic time change, we consider the deterministic time transformation $t \rightarrow t^{*}=t^{\alpha}$ in the normal diffusive process $X(t)$ (now in the physical time t), we obtain the SBM: $Y_{*}(t)=X(t^{*})$. Its equivalent Langevin equation is given by \cite{lim2002self,wu2008propagators,thiel2014scaled,Jeon2014ScaledBM}:
\begin{equation}
\dot{Y}_{*}(t)=\sqrt{2\,\alpha\,\sigma\,t^{\alpha - 1}}\,\xi(t),
\label{eq:1.6}
\end{equation}
with $\xi(t)$ being a white Gaussian noise (with the same properties as before, but in the physical time t). By using Eq.~\eqref{eq:1.6} we can prove straightforwardly that the MSD of $Y_{*}(t)$ is the same as Eq.~\eqref{eq:1.4} and that the corresponding FP equation is given by:
\begin{equation}
\derpar{}{t} P(y,t)= \alpha\,\sigma\,t^{\alpha - 1} \dersecpar{}{y}P(y,t), 
\label{eq:1.7}
\end{equation} 
which has time dependent diffusion coefficient \cite{batchelor1949diffusion}. This process preserves all the properties of Brownian motion \cite{lim2002self}: it is indeed Gaussian with time-dependent variance and Markov, as the monotonicity of the time change preserves the ordering of time. Furthermore, $Y_{*}(t)$ is self-similar and it has independent increments for non overlapping intervals. However, differently from Brownian motion, it is strongly non stationary \cite{Jeon2014ScaledBM}. Furthermore, $Y_{*}(t)$ turns out to be the mean-field approximation of the CTRW, as it describes the motion of a cloud of random walkers performing CTRW motion in the limit of a large number of walkers \cite{thiel2014scaled}. Recent investigation have also shown that SBM exhibits rich aging properties, which strongly differentiates it from the standard BM \cite{safdari2015aging}.

\subsection{\label{SubSec:1c}\textbf{ARBITRARY WAITING TIMES' DISTRIBUTION AND TIME TRANSFORMATIONS}}

In this section, we first focus on the generalization of Eqs.~(\ref{eq:1.1a}-\ref{eq:1.1b}) to arbitrary waiting time distributions of the underlying random walk \cite{chechkin2002retarding,chechkin2008generalized,magdziarz2009langevin,orzel2013accelerating,stanislavsky2014anomalous,Cairoli2014anomalous}. This extension is obtained naturally by choosing a different process $T(s)$ with the only assumption of it being non decreasing in order to preserve the causality of time. Thus, we consider $\eta(s)$ in Eq.~\eqref{eq:1.1b} to be an increasing L{\'e}vy noise with paths of finite variation and characteristic functional \cite{cont2013functional}: 
\begin{equation}
G[k(\tau)]=\Average{e^{\,-\int_{0}^{+\infty}k(\tau)\eta(\tau)\diff{\tau}}}=e^{\,-\int_{0}^{+\infty}\Phi(k(\tau))\diff{\tau}}.
\label{eq:1.8}
\end{equation}
Here $\Phi(k(\tau))$ is a non negative function with $\Phi(0)=0$ and strictly monotone first derivative, while $k(\tau)$ is a test function. We recall that for $\Phi(s)=s^{\alpha}$ we recover the CTRW model. Under these assumptions, the integrated process $T(s)$ is a a one-sided increasing L{\'e}vy process with finite variation. Furthermore, we assume $\eta(s)$ to be independent on the realizations of $\xi(s)$ in Eq.~\eqref{eq:1.1a}. As a consequence of the finite variation and the monotonicity of the paths of $T(s)$ respectively, $S(t)$ has continuous and monotone paths, with this second property implying the fundamental relation \cite{baule2005joint}: 
\begin{equation}
\Theta(s-S(t))=1-\Theta(t-T(s)).
\label{eq:1.9}
\end{equation}
Similarly to Eq.~\eqref{eq:1.3}, we can derive the corresponding MSD by recalling that $\widetilde{h}(s,\lambda)=\frac{\Phi\!\left(\lambda\right)}{\lambda} e^{-s \Phi\!\left(\lambda\right)}$ \cite{stanislavsky2014anomalous,Cairoli2014anomalous}:
\begin{equation}
\Average{Y^2(t)}=2\, \sigma \int_0^{t} K(\tau)\diff{\tau}, 
\label{eq:1.10}
\end{equation}
for $K(t)$ being related in Laplace space to $\Phi(s)$ by:  
\begin{equation}
\widetilde{K}(\lambda)=\frac{1}{\Phi(\lambda)}.
\label{eq:1.11}
\end{equation}
Furthermore, the PDF of $Y(t)$ is obtained by solving the generalized FP equation \cite{Cairoli2014anomalous}: 
\begin{equation}
\derpar{}{t}P(y,t)=\sigma\dersecpar{}{y}\derpar{}{t}\int_0^{t}\,K(t-\tau)\,P(y,\tau)\diff{\tau}, 
\label{eq:1.12}
\end{equation}
whose solution in this particular case can be found for general $\Phi(s)$ in Laplace space:
\begin{equation}
\widetilde{P}(y,\lambda)=\frac{1}{\lambda}\sqrt{\frac{\Phi(\lambda)}{2 \sigma}}e^{-\sqrt{\frac{\Phi(\lambda)}{2 \sigma}}\left| y \right|}.
\label{eq:1.13}
\end{equation}

\begin{figure}
\includegraphics[scale=0.2]{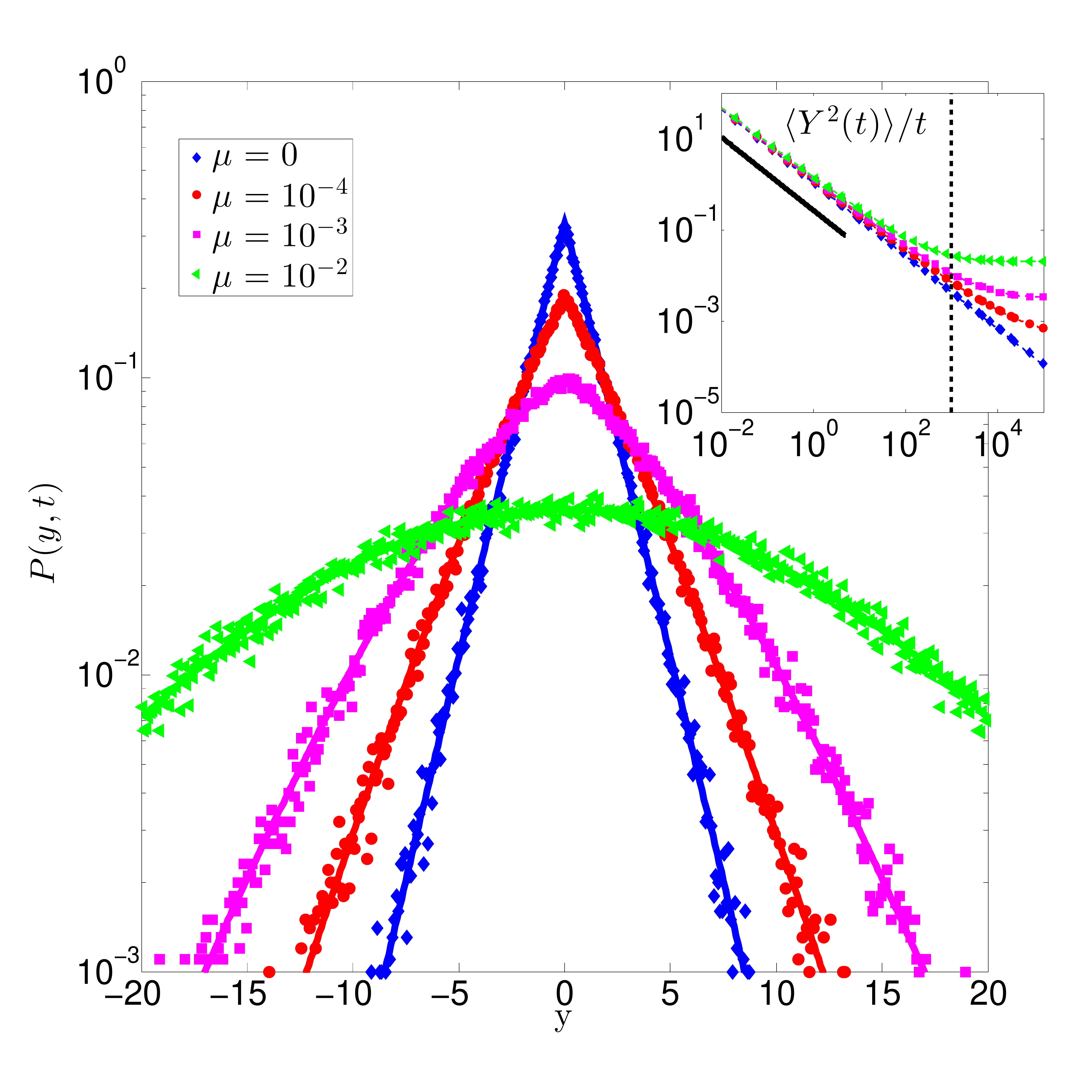}
\vspace{-0.6cm}
\caption{(Colors online) PDF (main) and MSD normalized to t (inset) of an anomalous process $Y(t)$ obtained by subordination of a pure diffusive process with a tempered stable L\'evy process of tempering index $\mu$ and stability parameter $\alpha=0.2$. The PDF is obtained by numerical Laplace inversion of Eq.~\eqref{eq:1.13} at $t=1000$ (black dotted lines in the inset) \cite{valko2005numerical}. The smooth transition from the non-Gaussian PDF typical of CTRWs ($\mu=0$) and the Gaussian one of normal diffusion ($\mu \rightarrow +\infty$) is evident, along with the corresponding transition from anomalous to normal scaling of the MSD for increasing $\mu$ at a fixed time. Simulations, obtained with the algorithms of \cite{kleinhans2007continuous,baeumer2010tempered}, agree perfectly with the analytical results.}  \label{fig:1}
\end{figure}

We look as an example at the case of a tempered stable L{\'e}vy noise with tempering index $\mu$ and stability index $\alpha$ \cite{Baeumer2002Stochastic}, which is obtained by setting $\Phi(\lambda)=(\mu+\lambda)^{\alpha}-\mu^{\alpha}$, i.e., $K(t)=e^{-\mu\,t}\,t^{\alpha-1}\,\MittagLeffler{\alpha}{\alpha}{(\mu\,t)^{\alpha}}$ \cite{janczura2011anomalous}. As already pointed out, the CTRW case is recovered by setting $\mu=0$, meaning that we do not truncate the long tails of the distribution, thus accounting for very long waiting times with a power-law decaying probability of occurrence. We plot in Figure~\ref{fig:1} the numerical Laplace inverse of Eq.~\eqref{eq:1.13} (main) and the corresponding MSD (inset) at a fixed time $t=1000$ (dotted line in the inset), which is given by \cite{mijena2014correlation}:
\begin{equation}
\Average{Y^2(t)}=\frac{2\,\sigma}{\mu^{\alpha}}\left[-1+\sum_{n=0}^{\infty} \frac{\gamma\!\left(\mu t;\alpha n\right)}{\Gamma(\alpha n)}\right],
\label{eq:1.14}
\end{equation}
with $\gamma(x;a)=\int_0^{x}e^{-t}t^{a-1}$ being an incomplete gamma function, leading to the asymptotic behaviour \cite{mijena2014correlation,Cairoli2014anomalous}: 
\begin{equation}
\label{eq:1.15}
\Average{Y^2(t)}\sim
\left\{
\begin{array}{cc}
\frac{2 \,\sigma}{\Gamma(1+\alpha)}t^{\alpha} & \quad t << 1 \\
\left(\frac{2\,\sigma}{\alpha} \mu^{1-\alpha} \right) t & \quad t>>1 
\end{array}
\right. 
\end{equation}
We remark that Eq.~\eqref{eq:1.15} does not apply to the long time scaling of CTRWs, for which it would predict a vanishing MSD. In fact, CTRWs do not exhibit a crossover from subdiffusive to normal behaviour, but their MSD scales as a power-law for all times. As expected, for $\mu=0$ we recover the typical non Gaussian shape of the PDF of a free diffusive CTRW \cite{metzler2000random}. However, for increasing values of $\mu$, the PDF of $Y(t)$, although still being non Gaussian, broadens, thus getting closer to a Gaussian. This has also evident consequences on the dynamical behaviour of the MSD, which for increasing values of $\mu$ goes from a pure subdiffusive scaling to a normal one (inset). 

We now discuss the corresponding extension of the SBM to arbitrary time transformations involving the kernel $K(t)$ obtained by Laplace inverse transform of Eq.~\eqref{eq:1.11}. We then generalize Eq.~\eqref{eq:1.6} by adopting $K(t)$ as the time dependent coefficient of the white noise:
\begin{equation}
\dot{Y}_{*}(t)=\sqrt{2 \sigma K(t)}\,\xi(t)=\zeta(t),
\label{eq:1.16}
\end{equation} 
where we define the correlated noise $\zeta(t)$ with $\Average{\zeta(t)}=0$ and two-point correlation function: $\Average{\zeta(t_1)\zeta(t_2)}=2 \sigma K(t_1)\delta(t_1-t_2)$. This explicit time dependence clearly signals that $\zeta(t)$ is a non stationary noise. It is easily shown that the MSD of $Y_{*}(t)$ is identical to the one of $Y(t)$ given by Eq.~\eqref{eq:1.10}. However, even if they share the same MSD, $Y(t)$ and $Y_{*}(t)$ provide different PDFs. Indeed, $Y_{*}(t)$ corresponds to a time rescaled Brownian motion $X(t^{*})$ with transformation:
\begin{equation}
t^{*}=\int_0^{t}K(\tau)\diff{\tau}. 
\label{eq:1.17}
\end{equation} 
In the case of the usual Brownian motion the corresponding diffusion equation has a Gaussian solution: $P(y,t)=\frac{1}{\sqrt{4 \pi \sigma t }}e^{-\frac{(y-y_0)^2}{4 \sigma t }}$ for the initial condition $P(y,0)=\delta(y-y_0)$. Since $Y_{*}(t)$ is just Brownian motion in the rescaled time $t^*$, we obtain similarly a Gaussian solution, provided we choose the same initial condition:  
\begin{equation}
P(y,t)=\frac{1}{\sqrt{4 \pi \sigma t^* }}e^{-\frac{(y-y_0)^2}{4 \sigma t^* }},
\label{eq:1.18}
\end{equation}
with $t^*$ as in Eq.~\eqref{eq:1.17}. We see that $P(y,t)$ is a solution of the diffusion equation:  
\begin{equation}
\derpar{}{t} P(y,t)= \sigma K(t) \dersecpar{}{y}P(y,t), 
\label{eq:1.19}
\end{equation}
with the time dependent diffusion constant: $D(t)=\sigma K(t)$. We remark that Eq.~\eqref{eq:1.6} can be recovered from these general results by setting $\Phi(\lambda)=\lambda^{\alpha}$, i.e., $K(t)=t^{\alpha - 1}/\Gamma(\alpha)$ and $t^{*}=t^{\alpha}/\Gamma(1+\alpha)$. However, in order to have exact equivalence, we need to neglect the constant multiplicative factors in both $K(t)$ and $t^*$ and make the following substitution: $\sigma \rightarrow \alpha\,\sigma$.           

\section{\label{Sec:2}\textbf{LANGEVIN FORMULATION OF ANOMALOUS PROCESSES IN PHYSICAL TIME}}
\subsection{\label{SubSec:2a}\textbf{DEFINITION OF THE NOISE}}

We proceed in this section to derive a Langevin description of the process $Y(t)$ defined in Eqs.~(\ref{eq:1.1a}-\ref{eq:1.1b}) directly in physical time. Starting from the explicit integral expression Eq.~\eqref{eq:add2} we can write the following: 
\begin{align}
Y(t)&=\sqrt{2\,\sigma}\int_0^{+\infty} \delta(s-S(t))\left[\int_0^s \xi(\tau)\diff{\tau}\right]\diff{s} \notag\\
&=\sqrt{2\,\sigma}\int_0^{+\infty} \left(-\derpar{}{s}\Theta(t-T(s))\right)\left[\int_0^s \xi(\tau)\diff{\tau}\right]\diff{s} \notag\\
&=\sqrt{2\,\sigma}\int_0^{+\infty}\Theta(t-T(s))\xi(s)\diff{s},
\label{eq:2.1}
\end{align}
where the fundamental relation of Eq.~\eqref{eq:1.9} is used to obtain the second equality and we then get the third one with an integration by parts. We remark that the boundary term $\left.\left[-\Theta(t-T(s))\int_0^{s}\xi(\tau)\diff{\tau}\right]\right|_{0}^{+\infty}$ is zero trivially for $s=0$, but it vanishes also for $s \rightarrow +\infty$ because $T(s)$ is increasing, thus always being bigger than any fixed (and finite) time t. Written as in Eq.~\eqref{eq:2.1}, $Y(t)$ is a differentiable (although in a generalized sense) function of time, so that we can take its derivative and obtain the equivalent Langevin equation:
\begin{equation}
\dot{Y}(t)=\sqrt{2\,\sigma}\,\overline{\xi}(t)
\label{eq:2.2}
\end{equation}
with the newly defined noise:
\begin{equation}
\overline{\xi}(t)=\int_0^{+\infty}\xi(s)\delta(t-T(s))\diff{s}, 
\label{eq:2.3}
\end{equation} 
whose properties are fully determined by the choice of the waiting times' distribution, i.e., equivalently of the function $\Phi(s)$ in Eq.~\eqref{eq:1.8}. If we recall the independence of $\xi(s)$ and $\eta(s)$, we can show that $\overline{\xi}(t)$ has zero average and two point correlation function: 
\begin{equation}
\Average{\overline{\xi}(t_1)\overline{\xi}(t_2)}= K(t_1)\delta(t_1-t_2),
\label{eq:2.4}
\end{equation} 
with $K(t)$ being specified by Eq.~\eqref{eq:1.11}. In Laplace space, indeed, $\Average{\overline{\xi}(t_1)\overline{\xi}(t_2)}$ is an integral of the two point characteristic function of $T(s)$, which can then be computed with Eq.~\eqref{eq:1.8}: $\Average{\widetilde{\overline{\xi}}(\lambda_1)\widetilde{\overline{\xi}}(\lambda_2)}=\frac{1}{\Phi\left(\lambda_1+\lambda_2\right)}$. By making its inverse Laplace transform, Eq.~\eqref{eq:2.4} follows straightforwardly. Consequently, the character of the noise $\overline{\xi}(t)$ significantly depends on the choice of the function $\Phi(s)$ in Eq.~\eqref{eq:1.8}. Thus, Eq.~\eqref{eq:2.2} defines a new Langevin model driven by a generalized and typically non Gaussian noise, except possibly for particular choices of the memory kernel $K(t)$, which is able to describe free diffusive anomalous processes with arbitrary waiting times distribution equivalently to the subordinated Eqs.~(\ref{eq:1.1a}-\ref{eq:1.1b}). 

We highlight that so far the standard renewal picture underlying conventional CTRWs still applies. Eq.~(\ref{eq:2.2}) is essentially the time derivative of Eq.~(\ref{eq:add2}) expressing the process in terms of stochastic increments. Differences will appear when external forces act on the diffusion processes. This case is discussed in Sec.~\ref{Sec:3}. 

\subsection{\label{SubSec:2b}\textbf{CHARACTERIZATION OF THE MULTIPOINT CORRELATION FUNCTIONS}}

The definition in Eq.~\eqref{eq:2.3} enables us to derive a complete characterization of the multipoint correlation structure of $\overline{\xi}(t)$. As a preliminary step, we derive the Laplace transform of the multipoint characteristic function of $T(s)$, i.e., $Z(t_1,s_1;\ldots;t_N,s_N)=\Average{\prod_{m=1}^{N} \delta(t_m-T(s_m))}$ $\forall N \in \mathbb{N}$. This is obtained by using the definition of Eq.~\eqref{eq:1.1b} as:    
\begin{equation}
\!\widetilde{Z}(\lambda_1,s_1;\ldots;\lambda_N,s_N)=\Average{\prod_{m=1}^{N} e^{-\lambda_m\int_0^{s_m}\eta(s^{\prime}_m)\diff{s^{\prime}_m}}}.\!\!
\label{eq:2.5}
\end{equation} 
Let us first assume an ordering for the sequence of times: $s_1<s_2<\ldots<s_N$ and compute the corresponding Eq.~\eqref{eq:2.5}. If we rearrange the exponent by separating successive intervals, we obtain:  
\begin{align}
&\widetilde{Z}(\lambda_1,s_1;\ldots;\lambda_N,s_N)= \notag \\
&=\Average{e^{-\lambda_N \int_{s_{N-1}}^{s_{N}}\eta(s^{\prime}_N)\diff{s^{\prime}_N}-\ldots-(\lambda_N+\ldots+\lambda_1)\int_{0}^{s_1}\eta(s^{\prime}_1)\diff{s^{\prime}_1} }} \notag\\
&=\Average{e^{-\sum_{m=0}^{N-1}\left[\left( \sum_{n=m+1}^{N}\lambda_n \right)\right] \int_{s_m}^{s_{m+1}}\eta(s^{\prime}_m)\diff{s^{\prime}_{m+1}}}} \notag\\
&=\prod_{m=0}^{N-1} e^{-(s_{m+1}-s_m)\Phi\left(\sum_{n=m+1}^{N}\lambda_n \right) },
\label{eq:2.6}
\end{align} 
where we define $s_0=0$ to simplify the notation and we exploited the independence of the increments of $T(s)$ to factorize the ensemble average. Furthermore, their stationariness together with Eq.~\eqref{eq:1.8} is then used to get Eq.~\eqref{eq:2.6}. However, in the general case where no a-priori ordering is assumed, we need to consider all the possible ordered sequences. We then introduce the group of permutations of N objects $S_N$, whose elements act on the sequence: $\bm{s}=(s_1,\ldots,s_N)$. When we make a permutation of $\bm{s}$, we obtain a new sequence with permuted indices: $\bm{s^{\prime}}=\left(s_{\sigma(1)},\ldots,s_{\sigma(N)}\right)$. All the possible orderings of $\bm{s}$ are thus obtained by summing over all the permutations in $S_N$. If we assume that $\sigma(s_0)=0$, $\forall \sigma \in S_{N}$, i.e., the initial time is kept fixed by the permutations, and we use the result of Eq.~\eqref{eq:2.6}, we derive:
\begin{multline}
\!\!\!\!\widetilde{Z}(\lambda_1,s_1;\ldots;\lambda_N,s_N)\!=\!\!\sum_{\sigma\in S_{N}}\prod_{m=0}^{N-1}\Theta\!\left(s_{\sigma(m+1)}-s_{\sigma(m)}\right) \times \\
\times e^{-\left(s_{\sigma(m+1)}-s_{\sigma(m)}\right) \Phi \left( \sum_{n=m+1}^{N}\lambda_{\sigma(n)} \right)} 
\label{eq:2.7}
\end{multline}
with the ordering of the permuted sequence being specified by the product of Heaviside functions. By factorizing out the first term, we obtain the fundamental result:
\begin{align}
&\widetilde{Z}(\lambda_1,s_1;\ldots;\lambda_N,s_N)\!=\!\!\sum_{\sigma\in S_{N}}\! e^{-s_{\sigma(1)}\Phi\left(\sum_{m=1}^{N}\lambda_{m}\right)}\times \label{eq:2.8}\\
&\!\!\times\!\! \prod_{m=1}^{N-1}\!\Theta\!\left(s_{\sigma(m+1)}\! - \! s_{\sigma(m)}\right) \!e^{-\left(s_{\sigma(m+1)}-s_{\sigma(m)}\!\right) \Phi\left( \sum_{n=m+1}^{N}\lambda_{\sigma(n)} \!\right)} \notag
\end{align}
As an example, we recover the two-point case: $\widetilde{Z}(\lambda_1,s_1;\lambda_2,s_2)$ \cite{Cairoli2014anomalous}. If we put $N=2$ in Eq.~\eqref{eq:2.8} and we consider the two possible permuted sequences: $\bm{s}=(s_1,s_2)$ and $\bm{s^{\prime}}=(s_2,s_1)$, we obtain:
\begin{multline}
\!\!\widetilde{Z}(\lambda_1,s_1;\lambda_2,s_2)\!=\!\Theta\!\left(s_2-s_1\right) e^{-s_1\Phi\left(\lambda_1+\lambda_2\right)}e^{-\left(s_2-s_1\right) \Phi\left(\lambda_2 \right)} \\
+\Theta\!\left(s_1-s_2\right) e^{-s_2\Phi\left(\lambda_1+\lambda_2\right)} e^{-\left(s_1-s_2\right) \Phi\left(\lambda_1 \right)}.
\label{eq:2.9}
\end{multline}

We now use Eq.~\eqref{eq:2.8} to compute the correlation functions of $\overline{\xi}(t)$. Indeed, we obtain from Eq.~\eqref{eq:2.3} $\forall N \in \mathbb{N}$:
\begin{multline}
\Average{\overline{\xi}(t_1)\ldots\overline{\xi}(t_{2N})}=\left[ \prod_{m=1}^{2 N}\int_0^{+\infty}\diff{s_m} \right] \times \\
\times \Average{\prod_{m=1}^{2 N}\xi(s_m)}\Average{\prod_{m=1}^{2 N}\delta(t_m-T(s_m))},
\label{eq:2.10}
\end{multline}
where the average is factorized due to the independence of the noises. If we recall the Wick theorem holding for the white noise $\xi(s)$ \cite{isserlis1918formula,wick1950evaluation}:
\begin{multline}
\Average{\prod_{j=1}^{2 N}\xi(t_j)}\!\!\!=\!\!\frac{1}{N 2^N}\!\!\sum_{\sigma \in S_{2 N}} \prod_{j=1}^{N}\!\Average{\xi\!\left(t_{\sigma(2 N - j + 1)}\right)\!\xi\!\left(t_{\sigma(j)}\right)} \\
=\frac{1}{N 2^N}\sum_{\sigma \in S_{2 N}} \prod_{j=1}^{N}\delta\!\left(t_{\sigma(2 N - j + 1)} - t_{\sigma(j)}\right)
\label{eq:2.11}
\end{multline}
and we substitute it in Eq.~\eqref{eq:2.10}, we can derive: 
\begin{align}
&\Average{\overline{\xi}(t_1)\ldots\overline{\xi}(t_{2 N})}=\frac{1}{N 2^N}\sum_{\sigma \in S_{2 N}}\left[ \prod_{m=1}^{2 N}\int_0^{+\infty}\diff{s_m}\right]\times \notag\\ 
&\,\times \prod_{j=1}^{N}\delta\!\left(s_{\sigma(2 N - j + 1)}-s_{\sigma(j)}\right)\Average{\prod_{i=1}^{2 N} \delta(t_i-T(s_i))} \notag\\ 
&=\frac{1}{N 2^N}\sum_{\sigma \in S_{2 N}} \left[\prod_{m=1}^{N}\int_0^{+\infty}\diff{s_{\sigma(m)}}\right] \times \label{eq:2.12}\\
&\,\times \Average{\prod_{j=1}^{N} \delta\!\left(t_{\sigma(2 N - j + 1)}-T\left(s_{\sigma(j)}\right)\right)\!\delta\!\left(t_{\sigma(j)}-T\left(s_{\sigma(j)}\right)\right)} \notag
\end{align}
with N integrals being solved by using the delta functions obtained from $\Average{\xi(s_1)\ldots\xi(s_{2 N})}$. If we make a Laplace transform of Eq.~\eqref{eq:2.12}, we obtain an expression involving $\widetilde{Z}(\lambda_1,s_1;\lambda_2,s_2;\ldots;\lambda_N,s_N)$:
\begin{multline}
\Average{\prod_{j=1}^{2 N}\widetilde{\overline{\xi}}(\lambda_j)}=\frac{1}{N 2^N}\sum_{\sigma \in S_{2 N}} \left[\prod_{m=1}^{N}\int_0^{+\infty}\diff{s_{m}}\right] \times \label{eq:2.13} \\
\, \times \widetilde{Z}\!\left(\lambda_{\sigma(1)}\!+\!\lambda_{\sigma(2 N)},s_1;\ldots;\lambda_{\sigma(N)}\!+\!\lambda_{\sigma(N+1)},s_N\right),
\end{multline}
which can thus be further simplified with Eq.~\eqref{eq:2.8}. By substitution and by making a further permutation of the indices, we obtain: 
\begin{align}
\!\!\!\!\!\!\!&\Average{\prod_{j=1}^{2 N}\widetilde{\overline{\xi}}(\lambda_j)}\!\!=\!\!\frac{1}{N 2^N}\!\!\!\sum_{\sigma \in S_{2 N}}\!\sum_{\sigma^{\prime} \in S_{N}}\!\!\left[\prod_{m=1}^{N}\int_0^{+\infty}\diff{s_{\sigma^{\prime}(m)}}\right]\!\!\times \notag\\
&\!\!\times e^{-s_{\sigma^{\prime}(1)}\Phi\left(\sum_{m=1}^{N}\lambda_{m}\!\right)}\prod_{m=1}^{N-1}\left[\Theta\left(s_{\sigma^{\prime}(m+1)}-s_{\sigma^{\prime}(m)}\right)\times\right. \label{eq:2.14} \\
&\!\!\left. \times e^{-\left(s_{\sigma^{\prime}(m+1)}-s_{\sigma^{\prime}(m)}\right) \Phi\left( \sum_{n=m+1}^{N}\left(\lambda_{\sigma(\sigma^{\prime}(n))}+\lambda_{\sigma(2N-\sigma^{\prime}(n)+1)}\right) \right)} \right], \notag
\end{align}
where the N integrals can then be solved by making suitable changes of variables. This leads to the following result for the Laplace transform of even multipoint functions of $\overline{\xi}(t)$: 
\begin{align}
&\Average{\widetilde{\overline{\xi}}(\lambda_1)\ldots\widetilde{\overline{\xi}}(\lambda_{2 N})}\!=\!\frac{1}{N 2^N \Phi\!\left(\sum_{m=1}^{2 N}\lambda_m \right)}\sum_{\sigma \in S_{2 N}}\!\times \label{eq:2.15} \\
&\times\!\! \sum_{\sigma^{\prime} \in S_{N}}\prod_{m=1}^{N-1}\frac{1}{\Phi\!\left(\sum_{n=m+1}^{N}\left(\lambda_{\sigma\left(\sigma^{\prime}(n)\right)}+\lambda_{\sigma\left(2N-\sigma^{\prime}(n)+1\right)}\right)\!\right)}. \notag
\end{align}
We remark that odd multipoint correlation functions are zero; indeed, if we make the substitution $2N \rightarrow 2N+1$ in Eq.~\eqref{eq:2.10}, we obtain an expression depending on the odd multipoint correlation functions: $\Average{\xi(s_1)\ldots\xi(s_{2N+1})}$, which vanish $\forall N \in \mathbb{N}$. The corresponding quantities in time are derived by making the inverse Laplace transform of Eq.~\eqref{eq:2.15}, which can be written as a $2N-$fold convolution: 
\begin{widetext}
\begin{subequations}
\begin{align}
\Average{\overline{\xi}(t_1)\ldots\overline{\xi}(t_{2N})}&=\frac{1}{N 2^N}K(t_1)\prod_{i=1}^{N-1}\delta(t_{i+1}-t_{i})*_{2N} g(t_1,t_2,\ldots,t_{2N-1},t_{2N})
\label{eq:2.16a} \\
g(t_1,t_2,\ldots,t_{2N-1},t_{2N})&=\LaplaceInverse{\sum_{\sigma \in S_{2 N}}\sum_{\sigma^{\prime} \in S_{N}}\prod_{m=1}^{N-1}\frac{1}{\Phi\left(\sum_{n=m+1}^{N}(\lambda_{\sigma(\sigma^{\prime}(n))}+\lambda_{\sigma(2N-\sigma^{\prime}(n)+1)})\right)}}
\label{eq:2.16b}
\end{align}
\end{subequations}
\end{widetext}
with $K(t)$ being the memory kernel defined in Eq.~\eqref{eq:1.11}. The set of Eqs.~(\ref{eq:2.16a}-\ref{eq:2.16b}) can then be used to compute all the multipoint correlation functions of $\overline{\xi}(t)$ and consequently of $Y(t)$. It is straightforward to recover the two point case of Eq.~\eqref{eq:2.4}, whereas we provide below as an example the four point function. First, we need to compute Eq.~\eqref{eq:2.16b} in time space:
\begin{multline}
g(t_1,t_2,t_3,t_4)=\left[K(t_1)\delta(t_2-t_1)\delta(t_3)\delta(t_4)\right. \\
\left. +K(t_1)\delta(t_1-t_3)\delta(t_2)\delta(t_4)+K(t_2)\delta(t_2-t_4)\delta(t_1)\delta(t_3) \right. \\
\left. +K(t_1)\delta(t_1-t_4)\delta(t_2)\delta(t_3)+K(t_2)\delta(t_2-t_3)\delta(t_1)\delta(t_4)\right. \\
\left. +K(t_3)\delta(t_3-t_4)\delta(t_1)\delta(t_2)\right],
\label{eq:2.17}
\end{multline}
and then solve the $\left. \frac{(2N)! N!}{N 2^N}\right|_{N=2}=6$ convolution integrals of Eq.~\eqref{eq:2.16a}. This can be done explicitly, so that we derive:  
\begin{align}
&\Average{\overline{\xi}(t_1)\overline{\xi}(t_{2})\overline{\xi}(t_3)\overline{\xi}(t_{4})}=\left[ K(\min(t_1,t_2)) \times \right. \notag\\ 
&\times K(|t_1-t_2|)\delta(t_4-t_1)\delta(t_3-t_2)+K(\min(t_1,t_3))\times \notag\\
&\times K(|t_1-t_3|)\delta(t_4-t_3)\delta(t_2-t_1)+K(\min(t_1,t_4))\times \notag\\
&\left. \times K(|t_1-t_4|)\delta(t_3-t_1)\delta(t_4-t_2) \right].
\label{eq:2.18}
\end{align}
We verified that the same similar structure of the time dependent coefficients is shared by the six point correlation function. Considering the recursive structure evident from Eqs.~(\ref{eq:2.16a}-\ref{eq:2.16b}), we conjecture the following formula for the even correlation functions in time space (with $t_{0}=0$ kept fixed by the permutations): 
\begin{multline}
\Average{\prod_{j=1}^{2 N} \overline{\xi}(t)}\!\!=\!\!\frac{1}{N 2^N}\!\! \sum_{\sigma \in S_{2 N}}\!\prod_{m=1}^{N}\! \delta\!\left(t_{\sigma(2 N - m + 1)}-t_{\sigma(m)}\right)\! \times\\
\times \sum_{\sigma^{\prime} \in S_{N}} \Theta\!\left(t_{\sigma(\sigma^{\prime}(m))}-t_{\sigma(\sigma^{\prime}(m-1))}\right)\times\\ 
\times K\!\left(t_{\sigma(\sigma^{\prime}(m))}-t_{\sigma(\sigma^{\prime}(m-1))}\right).
\label{eq:2.19}
\end{multline}

\subsection{\label{SubSec:2c}\textbf{COMPARISON WITH THE SCALED BM}}

Once the underlying noise structure of the CTRW is revealed by Eqs.~(\ref{eq:2.16a}-\ref{eq:2.16b}-\ref{eq:2.19}), a comparison with the corresponding multipoint correlation functions of the noise $\zeta(t)$ of the SBM reveals important common features of these two processes. Indeed, the correlation functions of $\zeta(t)$ are obtained straightforwardly by using the definition of Eq.~\eqref{eq:1.16} and the Wick theorem in Eq.~\eqref{eq:2.11}:  
\begin{multline}
\Average{\prod_{j=1}^{2 N}\zeta(t_j)}=\frac{1}{N 2^N}\!\!\sum_{\sigma \in S_{2N}}\prod_{m=1}^{N} K\!\left(t_{\sigma(m)}\right)\times \\
\times \delta\!\left(t_{\sigma(2 N - m + 1)}-t_{\sigma(m)}\right). 
\label{eq:2.20}
\end{multline}
Odd correlation functions of $\zeta(t)$ are zero as for $\overline{\xi}(t)$. As an example to better clarify our discussion, we provide the four point correlation function:
\begin{multline}
\!\!\!\!\!\!\Average{\zeta(t_1)\zeta(t_2)\zeta(t_3)\zeta(t_4)}\!=\! K(t_1) K(t_2) \delta(t_1-t_3)\delta(t_2-t_4) \\ 
+ K(t_1) K(t_3) \delta(t_1-t_2)\delta(t_3-t_4) \\
+ K(t_2) K(t_4) \delta(t_1-t_4)\delta(t_2-t_3).
\label{eq:2.21}
\end{multline}  
A first remark has to be done when we set $N=2$, thus studying the two point correlation function. Indeed, this is found to be the same for both the noises $\overline{\xi}(t)$ and $\zeta(t)$ and equal to Eq.~\eqref{eq:2.4}, thus explaining why the corresponding integrated processes $Y(t)$ and $Y_{*}(t)$ share the same MSD. On the contrary, differences are evident only if we look at the higher order correlation functions. Thus, the two integrated processes are distinguishable only by looking at quantities dependent on these higher order correlation functions, e.g., the PDFs or the corresponding higher order correlation functions of the integrated processes. Furthermore, by comparing Eqs.~(\ref{eq:2.19}-\ref{eq:2.20}), we can observe the same similar structure of the delta functions, typical of Gaussian processes, but with a different correlated and mainly not factorizable time structure of the coefficients in the case of $\overline{\xi}(t)$, which depends on the difference between successive time in the ordered sequences. This ultimately causes its non Gaussian typical character. In fact, in the specific case of a constant memory kernel, for all times or in some scaling limit, the two noises coincide and reduce to a standard Brownian motion.

\section{\label{Sec:3}\textbf{MODELS WITH EXTERNAL FORCES}}
We now consider models of anomalous processes in the presence of external forces \cite{metzler1998anomalous,metzler1999anomalousPRL,metzler1999anomalous,heinsalu2007use,magdziarz2008equivalence,eule2009subordinated}. Let us first focus on the random walk picture of the CTRW and assume that external force fields, which depends on the position of the walker, only modify its dynamics during the jumps. In the continuum limit, these forces are then naturally included in the Langevin equation of the process $X(s)$, thus modifying Eqs.~(\ref{eq:1.1a}-\ref{eq:1.1b}) into \cite{fogedby1994langevin}:
\begin{subequations}
\begin{align}
\dot{X}(s)&=F(X(s))+\sqrt{2\,\sigma}\xi(s), \label{eq:3.1a} \\
\dot{T}(s)&=\eta(s), \label{eq:3.1b}
\end{align}
\end{subequations} 
where the function $F(x)$ satisfies standard conditions \cite{revuz1999continuous}. However, different scenarios may be observed in experiments where forces can modify the position of the walker also during the waiting times between different jumps, without ultimately changing the underlying waiting times distribution. For instance, we would expect this situation to occur for the motion of an organelle inside the cytoplasm of a cell, which is freely migrating or driven by an external field. This different situation turns out not to be easily described with the time-change technique, as it is not clear how to modify the Langevin subordinated equations in order to take into account these further changes in the position variable. However, the characterization of the noise $\overline{\xi}(t)$ provided by Eqs.~(\ref{eq:2.16a}-\ref{eq:2.16b}), or equivalently by Eq.~\eqref{eq:2.19}, enables us to describe it by defining a new class of models, defined with  the Langevin equation:   
\begin{equation}
\dot{Y}(t)=F(Y(t))+\sqrt{2\,\sigma}\,\overline{\xi}(t). 
\label{eq:3.2}
\end{equation}

\begin{figure}
\includegraphics[scale=0.25]{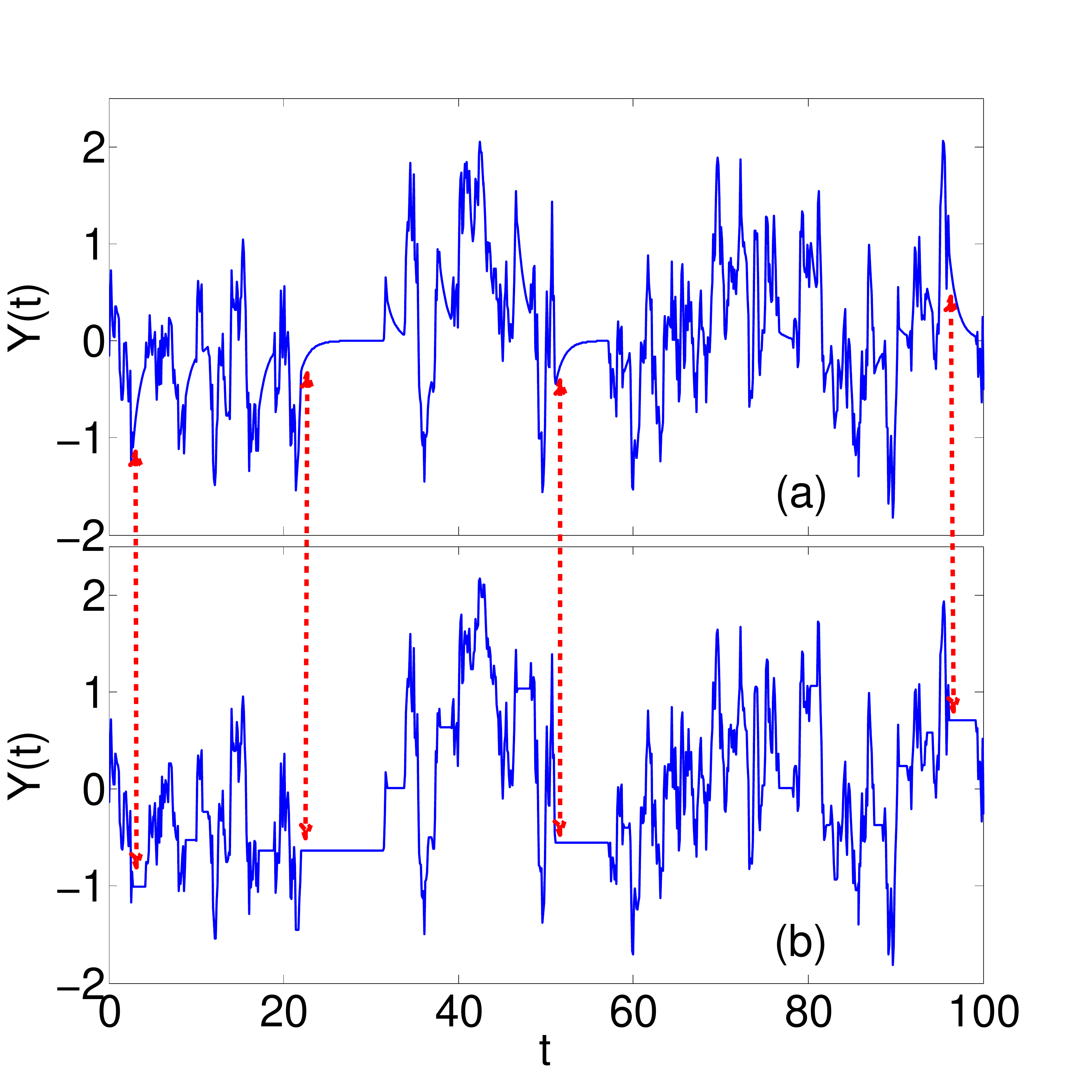}
\vspace{-0.6cm}
\caption{(Colors online) Simulated trajectories of a CTRW with a linear viscous-like force acting along its all time evolution (panel a, Eq.~\eqref{eq:3.2}) or acting only during the jumps (panel b, subordinated Eqs.~(\ref{eq:3.1a}-\ref{eq:3.1b})). Numerical algorithms are adapted from \cite{kleinhans2007continuous}. The difference on how the force affects the dynamics during trapping events is evident (dotted arrows): (a) the force acts on the particle, thus damping $Y(t)$ towards zero; (b) the force does not act, so that the particle gets physically stuck and $Y(t)$ is kept constant.}\label{fig:2}
\end{figure}

The difference between the dynamical behaviours generated by the two models becomes clear when we look at their simulated trajectories. In Figure~\ref{fig:2} we plot the paths of $Y(t)$ obtained both via subordination of Eqs.~(\ref{eq:3.1a}-\ref{eq:3.1b}) (panel b) and via integration of Eq.~\eqref{eq:3.2} (panel a) for a linear viscous-like force $F(x)=-\gamma x$ with $\gamma$ positive real constant. On the one hand, in the subordinated dynamics (dotted arrows, panel b) we observe time intervals where the corresponding anomalous process $Y(t)$ is constant, meaning that the walker, in the corresponding renewal picture, is waiting for the next jump to occur without any force being able to modify its position. On the other hand, during these same intervals the process $Y(t)$ generated by Eq.~\eqref{eq:3.2} is rapidly damped towards zero (dotted arrows, panel a), meaning that the walker is being driven by the external force. While indeed external forces act only during the jump times in the standard subordinated case, in our case they affect the dynamics of the system for all times, without intrinsically modifying the waiting times distribution and equivalently the relation between the number of steps and the physical time. We mention that another scenario involving external fields directly modifying the waiting time distribution of the random walk has been recently discussed in \cite{fedotov2014sub}, but this formalism does not have an evident connection with ours. 

Clearly, the inclusion of a force changes the renewal picture for the position variable $Y(t)$, which can no longer be expressed as a superposition of i.i.d. position increments as in Eqs.~(\ref{eq:add1},\ref{eq:add2}). These increments now depend on the accumulated position up to the time before the jump. However, the process $T(s)$, i.e., the stochastic process of the jump times parametrized by the arc-length still represents a renewal process, since the waiting times are unaffected by the force.

In the following, we present a comparison of the MSD obtained from Eqs.~(\ref{eq:3.1a}-\ref{eq:3.1b}-\ref{eq:3.2}) for a tempered stable subordinator as in Sec.~\ref{SubSec:1c} and for different choices of the external force $F(x)$. Except when explicitly stated we assume zero initial condition, so that the MSD coincides with the second order moment. We recall that the model of Eq.~\eqref{eq:3.2} defined with the time scaled noise $\zeta(t)$ instead of $\overline{\xi}(t)$ provides the same MSD.     

\subsection{\label{Sec:3a}\textbf{CONSTANT FORCE CASE}}

\begin{figure}[!htb]
\includegraphics[scale=0.22]{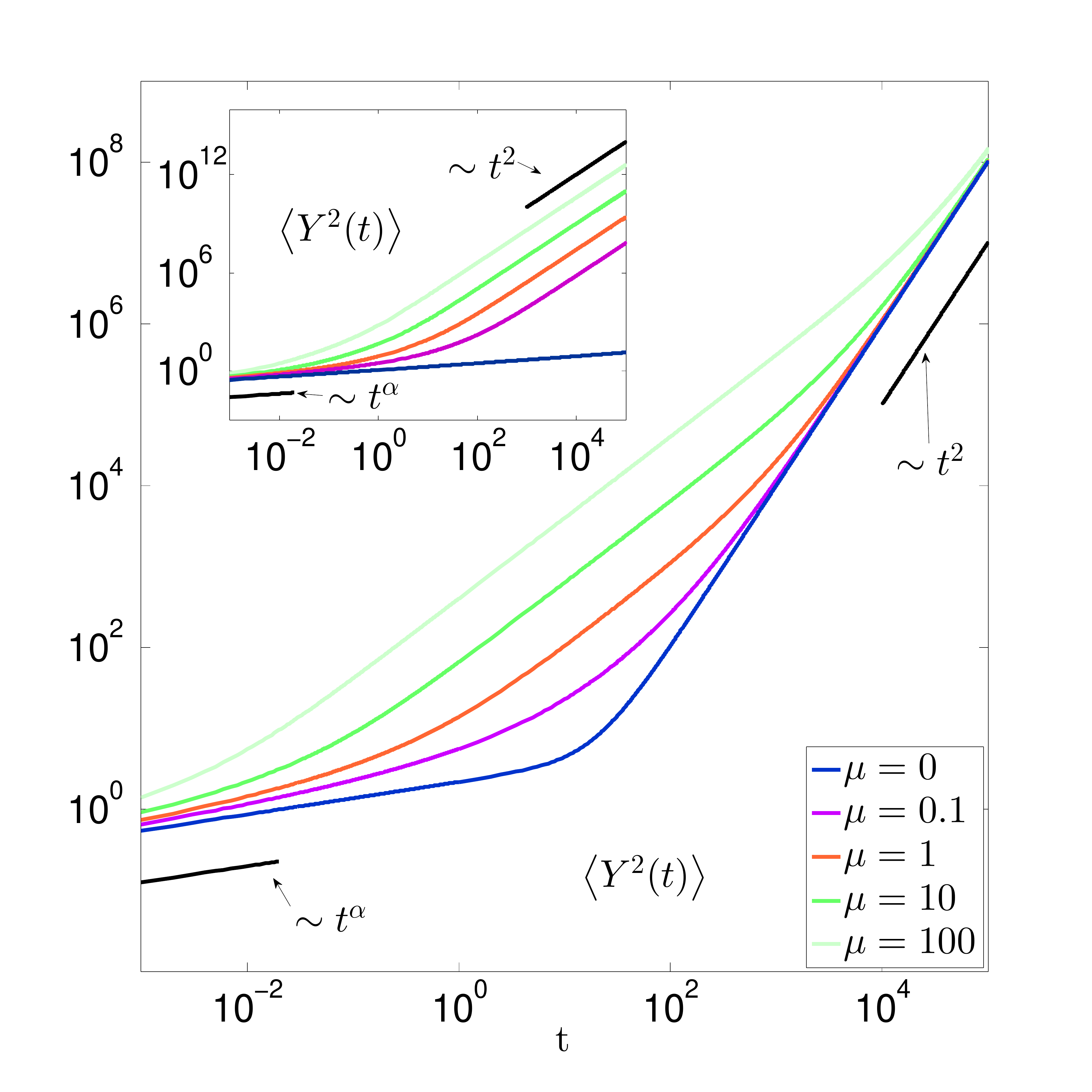}
\vspace{-0.8cm}
\caption{(Colors online) MSD of an anomalous process with tempered stable ($\alpha=0.2$) distributed waiting times in the presence of a constant force acting throughout the all dynamical evolution (main panel) or only during the jump times (inset). These two different scenarios are obtained with the $\overline{\xi}-$driven process or with the subordination technique, i.e., by numerical Laplace inverse transform of Eqs.~(\ref{eq:3.6}-\ref{eq:3.7}) respectively. The different long time scaling is evident: (main) the $\overline{\xi}-$driven process exhibits crossover to ballistic diffusion in all cases  and without any dependence on the tempering parameter $\mu$; (inset) the time-changed process exhibits crossover to ballistic diffusion with $\mu$-dependent scaling coefficient when $\mu \neq 0$, whereas it still scales as a power-law with exponent $2\,\alpha$ for $\mu \eq 0$.
}\label{fig:3}
\end{figure}

We first look at the case of a constant homogeneous force field: $F(Y(t))=F$ with $F \in \mathbb{R}^{+}$, for which Eq.~\eqref{eq:3.2} becomes:
\begin{equation}
\dot{Y}(t)=F+\sqrt{2 \sigma}\,\overline{\xi}(t).  
\label{eq:3.3}
\end{equation}
This equation can be solved formally for the exact trajectory of $Y(t)$:
\begin{equation}
Y(t)= F \, t + \sqrt{2 \sigma} \int_0^{t} \overline{\xi}(\tau)\diff{\tau}
\label{eq:3.4}
\end{equation}
and then used, together with Eq.~\eqref{eq:2.4}, to derive the MSD:
\begin{equation}
\Average{Y^2(t)}=F^2\,t^2+ 2\,\sigma \int_0^tK(\tau)\diff{\tau}
\label{eq:3.5}
\end{equation} 
or equivalently in Laplace transform as a function of $\Phi(s)$:
\begin{equation}
\Average{\widetilde{Y}^2(\lambda)}=\frac{2 F^2}{\lambda^3} + \frac{2 \sigma}{\lambda \Phi\left(\lambda\right)}.
\label{eq:3.6}
\end{equation}   
In the subordinated case, the MSD is computed with the same technique of Eq.~\eqref{eq:1.10} but with the different variance $\Average{X^2(s)}=\left(F^2\,s^2+2\,\sigma\,s \right)$. In Laplace space we obtain: 
\begin{equation}      
\Average{\widetilde{Y}^2(\lambda)}=\frac{2 F^2}{\lambda \left(\Phi(\lambda)\right)^2} + \frac{\sigma^2}{\lambda \Phi(\lambda)}.
\label{eq:3.7}
\end{equation}
The Laplace inverse transform of both Eqs.~(\ref{eq:3.6}-\ref{eq:3.7}) is plotted, together with their corresponding scaling behaviours, in Figure~\ref{fig:3} (main panel and inset respectively). In the small time limit, we find that both share the same power-law scaling of Eq.~\eqref{eq:1.15}. However, they differ between themselves and with Eq.~\eqref{eq:1.15} when we look at the scaling for long times. On the one hand, Eq.~\eqref{eq:3.6} provides the long time scaling: $\Average{Y^2(t)}\sim F^2\,t^2$. Hence, the constant force in this limit induces a crossover from subdiffusive to ballistic dynamics. Examples of this nonlinear behaviour have been recently discovered in the dynamics of chromosomal loci, which exhibit rapid ballistic excursions from their fundamental subdiffusive dynamics, caused by the viscoelastic properties of the cytoplasm \cite{weber2010bacterial,javer2014persistent}. Furthermore, it is evident that the exponential dumping of the waiting times' distribution does not affect the long time scaling, differently from the corresponding scaling of Eqs.~\eqref{eq:3.7}, which turns out to be (Figure~\ref{fig:3}, inset): 
\begin{equation}
\Average{Y^2(t)}\sim \left\{
\begin{array}{cc}
\left(\frac{F \mu^{1-\alpha}}{\alpha}\right)^2 t^2 & \quad \mu \neq 0 \\
\frac{2 F^2}{\Gamma(1+2 \alpha)}t^{2 \alpha} & \quad \mu=0 \\
\end{array}
\right.
\label{eq:3.8}
\end{equation} 
Thus, we find the same crossover to ballistic diffusion when $\mu \neq 0$, but with different $\mu$-dependent scaling coefficients, whereas in the CTRW case ($\mu=0$) this crossover pattern is lost and the power-law scaling is conserved, although with a different exponent.  

\subsection{\label{Sec:3b}\textbf{HARMONIC POTENTIAL CASE}}
\begin{figure}[!htb]
\includegraphics[scale=0.22]{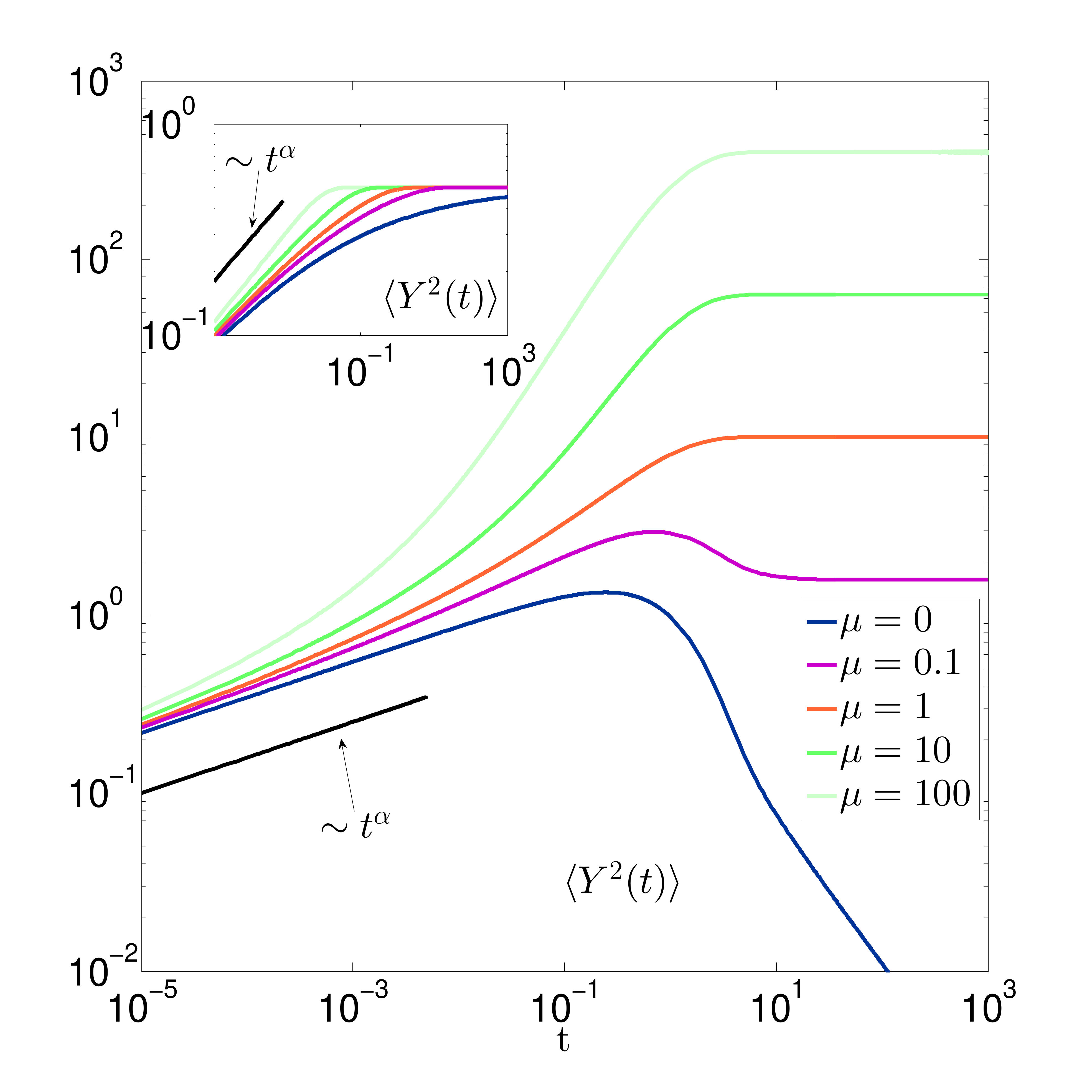}
\vspace{-0.8cm}
\caption{(Colors online) MSD of an anomalous process with tempered stable ($\alpha=0.2$) distributed waiting times in the presence of a linear viscous force acting at all times (main panel) or only during the jump times (inset). These two cases are obtained with the $\overline{\xi}-$driven process or with the subordination technique, i.e., by numerical Laplace inversion of Eqs.~(\ref{eq:3.10}-\ref{eq:3.11}) respectively. Whereas for small times the two processes exhibit subdiffusive scaling, their long time behaviour clearly differs: (main) the MSD of the $\overline{\xi}-$driven process decreases to zero in the CTRW case ($\mu=0$), whereas it converges to a $\mu$-dependent plateau for $\mu \neq 0$; (inset) in the subordinated case  all the curves converge to the same plateau.
}\label{fig:4}
\end{figure}

We now consider an external harmonic potential, leading to a friction-like force: $F(Y(t))=- \gamma Y(t)$ with $\gamma$ real positive constant. Thus, Eq.~\eqref{eq:3.2} provides the following:
\begin{equation}
\dot{Y}(t)=-\gamma Y(t) + \sqrt{2 \sigma}\,\overline{\xi}(t).  
\label{eq:3.9}
\end{equation}
As before, we can solve formally Eq.~\eqref{eq:3.9} for the trajectory of $Y(t)$ and use it together with Eq.~\eqref{eq:2.4} to compute the Laplace transform of the corresponding MSD:
\begin{equation}
\Average{\widetilde{Y}^2(\lambda)}=\frac{2 \sigma}{\left(\lambda+2\,\gamma\right)\Phi\left(\lambda\right)}. 
\label{eq:3.10}
\end{equation}
On the contrary, in the subordinated case we can proceed as in Eq.~\eqref{eq:1.10} by substituting: $\Average{X^2(s)}=\frac{\sigma}{\gamma}\left(1-e^{-2\,\gamma\,s}\right)$. One can thus obtain the result below:  
\begin{equation}
\Average{\widetilde{Y}^2(\lambda)}=\frac{\sigma}{\lambda\left[2 \gamma + \Phi(\lambda)\right]}.
\label{eq:3.11}
\end{equation} 
We plot in Figure~\ref{fig:4} the numerical inverse transform of Eqs.~(\ref{eq:3.10}-\ref{eq:3.11})  (main panel and inset respectively), along with their scaling behaviour for small times. While the small time scaling is in both cases the same as in Eq.~\eqref{eq:1.15}, we observe a different behaviour in the long time limit. Indeed, we find for Eq.~\eqref{eq:3.10} the following scaling laws: 
\begin{equation}
\Average{Y^2(t)}\sim \left\{
\begin{array}{cc}
\frac{\mu^{1-\alpha}}{\gamma \alpha} & \quad \mu \neq 0 \\
\frac{\sigma}{\gamma \Gamma(\alpha)}t^{\alpha-1} & \quad \mu=0 
\end{array}
\right.
\end{equation}
Thus, in the CTRW case the MSD decreases as a power-law towards zero. If we recall that this process is equal to the SBM up to the MSD, this is the same anomaly already reported in \cite{Jeon2014ScaledBM}. However, we also show that $Y(t)$ correctly converges to a plateau for $\mu \neq 0$, this being the expected dynamical behaviour of confined diffusion. By recalling that the waiting times are tempered stable distributed, the interpretation of the mentioned anomaly becomes clear. Indeed, the truncation of the power-law tails of the waiting times' distribution is fundamental to let the system find a stationary state, so that the MSD can converge to a plateau , which is typical of confined diffusion. In fact, no damping of the tails is done in the CTRW case, meaning that very long trapping events may still happen with non zero, but small probability. Thus, if we wait long enough, i.e., in the long time limit, these events eventually occur. However, Eq.~\eqref{eq:3.9} establishes that the system is affected by the external linear force also during such events, which then damps all the oscillations of the system. This clearly implies that the MSD should decrease to zero, because the system is not able to disperse and gets immobilized in $Y=0$. On the contrary, in the subordinated case the effect of the external force is stopped during the trapping events, so that the system does not get trapped in the zero position in the long time limit. Indeed, the MSD for different values of $\mu$ share the same long-time plateau: 
$\Average{Y^2(t)}\sim \frac{\sigma}{\gamma}.$

\section{\label{Sec:4}\textbf{CONCLUSION}}
In this work, we identified the underlying noise structure of a free diffusive CTRW with an arbitrary waiting times distribution and we defined its corresponding stochastic force. This enabled us to write a new Langevin equation, describing its dynamics directly in physical time and equivalently to the original formulation obtained with the subordination technique. We then derived a general formula, both in Laplace space and in physical time, providing all its multipoint correlation functions, which, although presenting the same time structure of Gaussian processes, have time dependent coefficients with a non factorizable dependence on the memory kernel generated by the corresponding subordinator of the equivalent time-changed formulation. Thus, except for the specific choice of a constant kernel, which recovers the factorizability of these coefficients, but reduces the noise to a standard Brownian motion, our new $\overline{\xi}-$noise was shown to be naturally both non Gaussian and non Markov. 

We then investigated the dynamics exhibited by processes driven by the $\overline{\xi}-$noise in the presence of external force fields and compared it with the one observed for usual subordinated processes. In general terms, we found that these processes belong to a new class of CTRW-like processes where external forces are exerted on the system at all times, i.e., both when the corresponding walker jumps or waits for the next jump to occur. Clearly, this is different from the original subordinated model, where external forces are implicitly assumed to modify the dynamics only during the jump times. Consequently, during the typical trapping events of subdiffusive dynamics the anomalous process $Y(t)$ becomes constant on the one hand, when it is generated via subordination, or it is deterministically driven by the force on the other hand, when it is driven by the $\overline{\xi}-$noise in physical time.

Furthermore, we found that these processes have the same MSD of those obtained with the characteristic noise of the SBM with time dependent diffusion coefficient being a function of their memory kernel. This relation indeed both provides a better interpretation for the anomaly reported in \cite{Jeon2014ScaledBM} and show that the correct scaling of the MSD typical of confined motion can be obtained by choosing more general kernels, which prevent an unbounded decay of the diffusion coefficient.

For future work it will be interesting to investigate the aging and ergodicity breaking properties of our new class of processes \cite{metzler2014anomalous}. This might further differentiate it from other anomalous processes such as the SBM. The properties of time-integrated observables of the $\overline{\xi}-$driven processes, which are expressed as functionals of their fluctuating trajectories, are also an open problem. For functionals of CTRWs, closed-form evolution equations can be derived that generalize the Feynman-Kac framework to anomalous processes \cite{Baule2006Damped,*Friedrich2006Exact,carmi2010distributions,*carmi2011fractional}. A further generalization to anomalous processes with arbitrary waiting time distributions has recently been obtained \cite{Cairoli2014anomalous}, which highlights the connection between the waiting time distribution and the memory kernel appearing in the fractional evolution equations. It will be highly interesting to investigate whether similar closed form equations can be formulated for functionals of trajectories driven by our $\overline{\xi}-$noise.

\begin{acknowledgements}          
This research utilized Queen Mary's MidPlus computational facilities, supported by QMUL Research-IT and funded by EPSRC grant EP/K000128/1.
\end{acknowledgements}

%

\end{document}